\documentclass[10pt,letterpaper]{article}
\usepackage{opex3}

\begin{document}

\title{Spectral dynamics of THz pulses generated by two-color laser filaments in air: The role of Kerr nonlinearities and pump wavelength}

\author{A. Nguyen$^{1,*}$, P. Gonz{\'a}lez de Alaiza Mart{\'i}nez$^1$, J. D{\'e}chard$^1$, I. Thiele$^2$, I. Babushkin$^3$, S. Skupin$^2$, and L. Berg\'e$^{1}$}

\address{
$^1$ CEA-DAM, DIF, 91297 Arpajon, France\\
$^2$ Univ.~Bordeaux - CNRS - CEA, Centre Lasers Intenses et Applications, UMR 5107, 33405 Talence, France\\
$^3$ Institute of Quantum Optics, Leibniz University Hannover, Welfengarten 1 30167, Hannover, Germany\\
$^*$ Corresponding author 
}

\email{alisee.nguyen@cea.fr}

\begin{abstract} 
We theoretically and numerically study the influence of both instantaneous and Raman-delayed Kerr nonlinearities as well as a long-wavelength pump in the terahertz (THz) emissions produced by two-color femtosecond filaments in air. Although the Raman-delayed nonlinearity induced by air molecules weakens THz generation, four-wave mixing is found to impact the THz spectra accumulated upon propagation via self-, cross-phase modulations and self-steepening. Besides, using the local current theory, we show that the scaling of laser-to-THz conversion efficiency with the fundamental laser wavelength strongly depends on the relative phase between the two colors, the pulse duration and shape, rendering a universal scaling law impossible. Scaling laws in powers of the pump wavelength may only provide a rough estimate of the increase in the THz yield. We confront these results with comprehensive numerical simulations of strongly focused pulses and of filaments propagating over meter-range distances.
\end{abstract}



\section{Introduction}

Laser filaments produced by ultrashort light pulses proceed from the dynamic balance between Kerr self-focusing and plasma generation \cite{Chin:cjp:83:863,Berge:rpp:70:1633}. The interplay of these nonlinear effects contributes to broaden the pulse spectrum, promote self-compression \cite{Berge:prl:100:113902} and self-guide intense optical wave packets over remote distances for, e.g., sensing applications \cite{Kasparian:sc:301:61}. Filamentation of pulses with different frequencies has recently been proposed as an innovative way to downconvert optical radiation into the THz range \cite{Wang:apl:95:131108} and create broadband THz sources remotely \cite{Liu:np:4:627}. Terahertz pulse generation by two-color (fundamental and its second harmonic) laser filaments mainly results from photocurrents \cite{Kim:oe:15:4577,Kim:np:2:605}, through which photoionization in tunneling regime drives the conversion process \cite{Thomson:lpr:1:349}. However, because THz radiation is also emitted by optical rectification through four-wave mixing \cite{Cook:ol:25:1210}, Kerr self-focusing can contribute to the overall THz yield \cite{Borodin:ol:38:1906}. Kerr-driven THz sources are expected to emit on-axis, mostly during the early self-focusing stage preceding ionization of air molecules \cite{Andreeva:prl:116:063902}, and their characteristic spectrum exhibits a parabolic distribution being maximum at non-zero frequency \cite{Cook:ol:25:1210}. At clamping intensities $>50$ TW/cm$^2$ from which Kerr self-focusing is stopped by plasma defocusing, THz emission is dominated by photocurrents and the pulse spectrum peaks at smaller THz frequencies \cite{Berge:prl:110:073901}.\\

Several important issues still need to be addressed in this physics, such as the role of the Raman-delayed part of the Kerr nonlinearity, which arises due to the excitation of rotational and vibrational transitions of the molecular constituents of air \cite{Penano:pre:68:056502}. Another point is the impact of longer pump wavelengths $(\lambda)$, in particular in the near-infrared domain $\sim 1.6$ $\mu$m that may be preferred for, e.g., ocular safety reasons. Antecedent studies \cite{Berge:prl:110:073901,Clerici:prl:110:253901} reported impressive growths of THz energy yields scaling as $\lambda^{\alpha}$ with $\alpha \simeq 4-5$. More recent ones \cite{Debayle:oe:22:13691,Debayle:pra:91:041801}, however, showed that, although single-color laser pulses produce THz energy yields increasing like $\lambda^4$, no similar conclusion could be inferred for two-color pulses.\\

The present paper aims at clarifying the role of the pump wavelength and of the Kerr nonlinearities (instantaneous and delayed) in filament-driven THz pulse generation. We display numerical evidence that four-wave mixing impacts the THz generation process over long propagation distances, even in intensity regimes where the photocurrent mechanism is the dominating THz emitter. We also demonstrate that the Raman-delayed Kerr nonlinearity does not contribute as a THz source. By means of the local current (LC) model \cite{Babushkin:njp:13:123029}, we moreover explain the variations in the THz field strength with respect to the fundamental wavelength of the optical radiation. In this respect we emphasize the role of the electron current component associated with the high-frequency laser pulse, whose fundamental at longer wavelength affects more significantly the THz spectrum. Importantly, the scaling of the THz field strength with the fundamental pump wavelength is shown to vary with the relative phase between the fundamental and second harmonic, the pulse envelope and duration, so that no universal $\lambda$-dependent scaling is achievable with a two-color pulse. Despite this, the conversion efficiencies reported from the LC model show that the THz yield is roughly scaling as $\lambda^{\alpha}$ with $\alpha > 4$ for small relative phases between the two colors and $\alpha > 2$ on the average. For focused pulses \cite{Clerici:prl:110:253901}, these results are confirmed by direct simulations employing a unidirectional pulse propagator \cite{Kolesik:prl:89:283902,Kolesik:pre:70:036604}. Smaller gain factors are achieved by meter-range two-color filaments due to the generation of weaker plasma densities.\\

The paper is organized as follows: Section \ref{S2} proposes a one-dimensional (1D) approach combining known laser-driven THz sources. It recalls that, in the range of intensities reached by two-color filaments in air, photoionization and to a lesser extent the Kerr nonlinearity are the principal players in THz generation. Section \ref{S3} discusses analytical estimates of the laser-to-THz conversion efficiency when taking a Raman-delayed nonlinearity into account and when increasing the fundamental laser wavelength. Section \ref{S4} verifies our analytical statements through three-dimensional (3D) comprehensive numerical simulations for both focused and filamentary pulses.

\section{Transverse versus longitudinal THz fields - A 1D Approach} \label{S2}

Before proceeding with full 3D simulations, we find it instructive to examine the dynamics of THz fields produced in-situ, i.e., inside the plasma channel created by two-color filaments in air. For simplicity, we use a reduced model discriminating THz transverse $(x)$ from longitudinal $(z)$ currents for a laser pulse polarized along the $x$-axis. Discarding ion motions, the current density induced by the photoionized electrons is ${\vec J} \simeq -e N_e {\vec v}_e$, where $N_e$ and ${\vec v}_e$ are the free electron density and velocity. Following Sprangle {\it et al.} \cite{Sprangle:pre:69:066415}, the electron current obeys the following equation set in the non-relativistic interaction regime:
\begin{equation}
\label{e3}
(\partial_t + \nu_c ) {\vec J} = \frac{e^2}{m_e} N_e {\vec E} + {\vec \Pi},
\end{equation}
\begin{equation}
\label{e4}
{\vec \Pi} = - \frac{e}{m_e} {\vec J} \times {\vec B} + ({\vec J} \cdot {\vec \nabla}) \left(\frac{ {\vec J}}{e N_e}\right) + \frac{{\vec J}}{e N_e} ({\vec \nabla} \cdot {\vec J}),
\end{equation}
where $\nu_c$ is the electron collision rate equal to 10 ps$^{-1}$, ${\vec B}$ is the magnetic field associated with the electric field ${\vec E}$ that includes both the laser field $E_L {\vec e}_x$ and secondary (THz) radiation such as ${\vec {\tilde E}} = {\tilde E}_x {\vec e}_x + {\tilde E}_z {\vec e}_z$. Assuming singly-ionized gases at moderate intensities $I_0 \leq 10^{15}$~W/cm$^2$, the growth of electron density is governed over femtosecond time scales by 
\begin{equation}
\label{Ne}
\partial_t N_e = W(E)(N_a - N_e),
\end{equation}
where $N_a$ is the initial gas density and $W(E)$ is a field-dependent ionization rate, for instance the quasi-static tunnel (QST) rate \cite{Landau:QM:65}
\begin{equation}
\label{QST}
W\left[E(t)\right]=\frac{4 (U_i/U_H)^{\frac{5}{2}} \nu_a}{|E(t)/E_a|}\exp{\left(-\frac{2(U_i/U_H)^{\frac{3}{2}}}{3|E(t)/E_a|}\right)},
\end{equation}
with $U_i$ being the ionization energy of the gas, $\nu_a = 4.13 \times 10^{16}$ Hz, $E_a = 5.14 \times 10^{11}$ V/m and $U_H = 13.6$ eV is the ionization potential of hydrogen. For air composed of $80\%$ of N$_2$ and $20\%$ of O$_2$, we shall start by considering ionization of oxygen molecules only, as their ionization potential ($12.1$ eV) is lower than that of nitrogen ($15.6$ eV). Photoionization of nitrogen molecules will be addressed later when considering, e.g., a field-dependent version of the cycle-averaged ionization rate derived by Perelomov, Popov and Terent'ev (PPT) \cite{Perelomov:spjetp:23:924}.\\

Equations (\ref{e3}) and (\ref{e4}), combined with Maxwell-Amp{\`e}re equation ${\vec \nabla} \times {\vec B} = c^{-2} \partial_t {\vec E} + \mu_0 {\vec J}$, readily provide the propagation equation for ${\vec E}$:
\begin{equation}
\label{EqEvec}
\left(\partial_t^2 + c^2 {\vec \nabla} \times {\vec \nabla} \times + \omega_{\rm pe}^2\right) {\vec E} + \nu_c \left(\partial_t {\vec E} +  c^2\int_{-\infty}^t {\vec \nabla} \times {\vec \nabla} \times {\vec E}(t') dt'\right) = - \frac{{\vec \Pi}}{\varepsilon_0},
\end{equation}
where $\omega_{\rm pe} = \sqrt{e^2 N_e/\varepsilon_0 m_e}$ is the electron plasma frequency ($\varepsilon_0 = 1/\mu_0 c^2$). We here neglect loss currents due to photoionization, which are small for our pump pulse configurations and in gas-based plasmas.\\

For technical convenience, Eq. (\ref{EqEvec}) is reduced to a one-dimensional, $z$-propagating model. Discarding temporarily the linear (chromatic) and nonlinear polarizations of air molecules, we omit the diffraction operators ($\partial_x = \partial_y = 0$), yielding
\begin{equation}
\label{EqPedrox}
\nu_c (\partial_t^2 - c^2 \partial_z^2)\int_{-\infty}^t E_x(t') \,dt' + \left(\partial_t^2 - c^2 \partial_z^2 + \omega_{\rm pe}^2 \right) E_x = - \frac{{\Pi_x}}{\varepsilon_0},
\end{equation}
\begin{equation}
\label{EqPedroz}
\left(\partial_t^2 + \nu_c \partial_t + \omega_{\rm pe}^2 \right) {\tilde E}_z = - \frac{{\Pi_z}}{\varepsilon_0}.
\end{equation}

As recently derived in \cite{Gonzalez:sr:2016}, further approximations can be applied such as assuming a laser field propagating with the sole variable $(z-ct)$. For the intensity range $<10^{15}$ W/cm$^{-2}$, we can furthermore neglect the longitudinal current component compared to its transverse counterpart that obeys $(\partial_t + \nu_c ) J_x = e^2 N_eE_x/m_e$ and use $B_y = - \partial_z \int_{-\infty}^t E_x (t')dt' \approx E_x/c$ for $B_y$ travelling like the laser pulse. By splitting $J_x = J_L + {\tilde J}_x$ on the expansion $E_x = E_L + {\tilde E}_x$, Eq. (\ref{EqPedrox}) reads as
\begin{equation}
\label{e6}
\left[ \partial_t^2 - c^2 \partial_z^2 + \partial_t (\partial_t + \nu_c)^{-1} \omega_{\rm pe}^2 \right] {\tilde E}_x = - \frac{1}{\varepsilon_0}\partial_t J_L.
\end{equation}
The equation for the longitudinal radiated field ${\tilde E}_z$ is easily expressed as
\begin{equation}
\label{e7}
(\partial_t^2 + \nu_c \partial_t + \omega_{\rm pe}^2) {\tilde E}_z =  \frac{e}{m_e \varepsilon_0 c} J_L E_L.
\end{equation}
Equations (\ref{e6}) and (\ref{e7}) both involve the current density computed on the laser field, $J_L = \varepsilon_0 (\partial_t + \nu_c)^{-1}\omega_{\rm pe}^2 E_L$. The transverse radiated field ${\tilde E}_x$ mainly proceeds from this current density \cite{Debayle:oe:22:13691,Babushkin:njp:13:123029}. With two colors, the product in $\partial_t J_L$ [Eq. (\ref{e6})] between the steplike increase of $N_e(t)$ and the fast oscillations of $E_L(t)$ acts as an efficient converter to low frequencies. By comparison, Eq.~(\ref{e7}) describes longitudinal plasma oscillations that develop over longer time scales after the laser field has interacted with the gas, as previously established in \cite{Sprangle:pre:69:066415,Damico:njp:10:013015}.\\

Including the optical polarization of a noble gas is straightforward by replacing into the Maxwell-Amp{\`e}re equation the electric field ${\vec E}$ by the displacement vector ${\vec D} = \epsilon_0 {\vec E} + {\vec P}_{\rm L} + {\vec P}_{\rm NL}$, where ${\vec P}_{\rm L}$ and ${\vec P}_{\rm NL}$ refer to linear and nonlinear polarization, respectively. In scalar description, this amounts to adding into the right-hand side of Eq. (\ref{EqPedrox}) the term $ - \varepsilon_0^{-1} \partial_t^2 (P_{\rm L} + P_{\rm NL})$. The Fourier transform of $P_{\rm L}$ involves the first-order frequency-dependent susceptibility $\chi^{(1)}(\omega)$ entering the optical linear index $n(\omega) = [1 + \chi^{(1)}(\omega)]^{1/2}$. $P_{\rm NL}$ involves the third-order susceptibility $\chi^{(3)}$ responsible for four-wave mixing and Kerr self-focusing (in multidimensional media). For molecular gases, the Kerr response admits a fraction $x_k$ ($0 \leq x_k \leq 1$) of delayed contribution due to Raman scattering by rotational molecular transitions. Assuming that the laser is not resonant with the transition frequencies \cite{Penano:pre:68:056502}, stimulated Raman scattering usually affects the total time-dependent refraction index of the medium. Its corresponding polarization component describes a delayed-Kerr nonlinearity. Written with the real electric field $E_x$ \cite{Berge:rpp:70:1633}, the overall nonlinear polarization can be expressed as
\begin{equation}
\label{kerrdel}
P_{\rm NL}(t) = (1 - x_k) \epsilon_0 \chi^{(3)} E_x^3(t) + x_k \frac{3}{2} \epsilon_0 \chi^{(3)} (h*E_x^2)(t) E_x(t),
\end{equation}
where
\begin{equation}
\label{h}
h(t) = \Theta(t) \frac{\tau_1^2+\tau_2^2}{\tau_1 \tau_2^2} \sin{(t/\tau_1)} \mbox{e}^{-t/\tau_2}.
\end{equation}
Here, symbol * means convolution product in time, $\Theta(t)$ is the usual Heaviside step function, $\tau_1$ and $\tau_2$ represent the rotational Raman time and dipole dephasing time, respectively. In air those quantities take the values $\tau_1 \approx 62$ fs and $\tau_2 \approx 77$ fs for the fraction $x_k = 0.5$ \cite{Penano:pre:68:056502,Pitts:josab:21:2008}. We shall assume that these values still hold for pump wavelengths up to 2 $\mu$m.\\

\begin{figure}
\centering \includegraphics[width=\columnwidth]{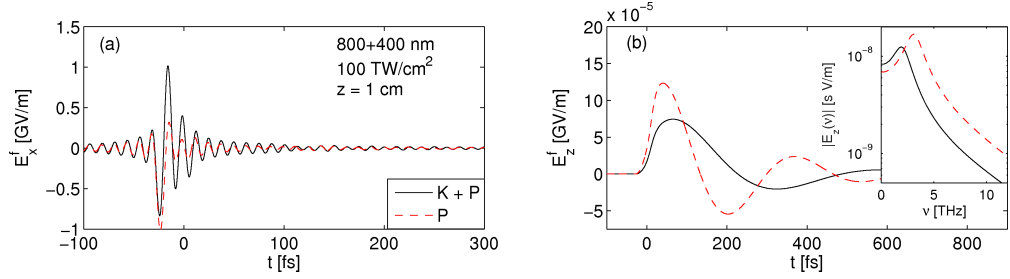}
\caption{(a) Transverse and (b) longitudinal fields computed at $z = 1$ cm from Eqs.~(\ref{EqPedrox}) and (\ref{EqPedroz}), and filtered in a 80-THz window for a 800+400-nm pulse with $I_0 = 100$~TW/cm$^2$, $\varphi =\pi/2$ and $r = 15\%$. The red dashed curve displays the THz field computed with the plasma response only (P); the solid curve corresponds to both Kerr (instantaneous) and plasma responses (K+P). Inset in (b) details the THz spectrum of the longitudinal field.}
\label{Fig1}
\end{figure}

\begin{figure}
\centering \includegraphics[width=\columnwidth]{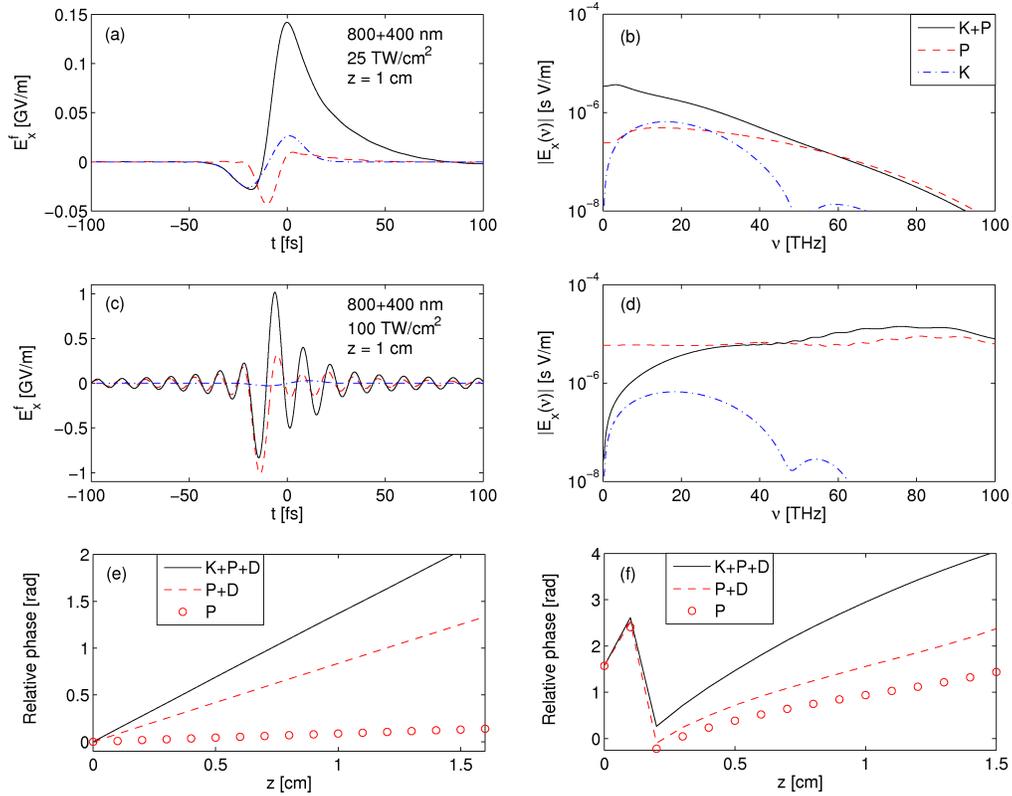}
\caption{(a) Transverse THz field for in-phase two-color 800+400-nm pulses ($r = 15\%$) at $z = 1$ cm by photoionization alone (P, red dashed curves), the Kerr response alone (K, blue dash-dotted curves) and both nonlinearities (K+P, black solid curves) at 25~TW/cm$^2$ with $\varphi=0$. (b) Corresponding spectra. (c) Transverse THz field at $z = 1$~cm by two-color pulses with 100 TW/cm$^2$ intensity and $\varphi = \pi/2$. These fields are filtered in the frequency window $\nu < 80$ THz. (d) Corresponding spectra. (e,f) Evolution of the relative phase $\varphi$ versus $z$ in the forward component of the electric field for (e) $I_0 = 25$ TW/cm$^2$ and (f) $I_0 = 100$~TW/cm$^2$, including or not linear dispersion (D, see legend). The phase jump near $z = 1$~mm is due to sharp distortions of the pulse profile induced by plasma generation.}
\label{Fig2}
\end{figure}

To start with, we integrate the (1+1)-dimensional Eqs. (\ref{EqPedrox}) and (\ref{EqPedroz}) using the initial condition 
\begin{equation}
\label{e2}
E_L(t,z=0) = \sqrt{\frac{2 I_0}{c\varepsilon_0}} \left[\sqrt{1-r} \,\mbox{e}^{- 2 \ln{2} \left(\frac{t}{\tau_p}\right)^2} \cos (\omega_0 t ) + \sqrt{r} \,\mbox{e}^{- 8 \ln{2} \left(\frac{t}{\tau_p}\right)^2}\cos (2 \omega_0 t + \varphi)\right],
\end{equation}
where $I_0$ is the mean pump intensity, $r$ is the relative intensity ratio of the second harmonic, $\varphi$ is the relative phase between the fundamental pulse with carrier frequency $\omega_0$ and its second harmonic. For $I_0 = 100$ TW/cm$^2$ and $r = 0.15$, Figs. \ref{Fig1}(a,b) show the transverse and longitudinal secondary fields filtered in a 80-THz frequency window for a two-color pulse propagating in air at ambient pressure. The full width at half maximum (FWHM) duration of the pump pulse centered at 800-nm wavelength is $\tau_p=40$ fs with half-duration for the $2\omega_0$ component. Simulations are performed using the quasi-static tunneling ionization, accounting or not for the Kerr nonlinearity, which we here consider instantaneous $(x_k=0)$ with a nonlinear index equal to $10^{-19}$~cm$^2$/W. For simplicity we first ignore linear dispersion whose action will be examined in Figs. \ref{Fig2}(e,f). At such intensities O$_2$ molecules undergo most of the ionization events for a neutral density $N_a = 5.4 \times 10^{18}$ cm$^{-3}$. With two-color pulses, the transverse field at the distance $z = 1$ cm is found to be orders of magnitude larger than the longitudinal one [Fig.~\ref{Fig1}(a)]. Plasma wakefield effects characterize the longitudinal field [Fig.~\ref{Fig1}(b)], with a long plasma wave formed behind the pulse head and oscillating at plasma frequency, as shown in the spectrum (see inset). Note that the plasma frequency varies due to small changes in $N_e$ caused by Kerr-induced self-steepening. At larger propagation distances, the transverse THz field increases, whereas the longitudinal field decreases even more (not shown). The Kerr response, although of minor role in the conversion process, increases the peak of the transverse THz field to some extent. Hence, at air-based filament intensities $\sim 100$ TW/cm$^2$, only the transverse secondary fields generated through photocurrents and four-wave mixing appear to be relevant in a two-color pulse configuration (see also \cite{Gonzalez:sr:2016}).\\

Figures \ref{Fig2}(a-d) detail the influence of the Kerr (blue dash-dotted curves), plasma (red dashed curves) and combined Kerr-plasma effects (black solid curves) when solving Eq. (\ref{EqPedrox}) for a fundamental pulse at 800 nm coupled with its second harmonic interacting with ambient air over short propagation distance ($z = 1$ cm). The left-hand side (LHS) column depicts the transverse THz field distributions computed from an inverse Fourier transform of the whole field within the frequency window $\nu < 80$ THz. The right-hand side (RHS) column shows corresponding spectra. At low intensities (25 TW/cm$^2$), which are characteristic of a self-focusing regime prior to ionization, the THz field exhibits a temporal profile shaped by the four-wave mixing contribution, here optimized with an initial zero relative phase between the two colors \cite{Cook:ol:25:1210}. However, the field maximum reaches much higher values in the presence of photoionization [Fig. \ref{Fig2}(a)]. The same conclusion holds in the reciprocal configuration, when the laser pulse evolves in self-channeling regime favoring THz emissions by photocurrents with $\varphi=\pi/2$ \cite{Kim:oe:15:4577,Li:apl:101:161104}. The THz field profile follows that created by a plasma source alone, but it appears noticeably amplified by the Kerr term [Fig.~\ref{Fig2}(c)], which also agrees with Ref. \cite{Andreeva:prl:116:063902}. Along propagation this optical nonlinearity contributes through self-, cross-phase modulations and self-steepening to broaden the pump and low-order (second as well as third) harmonics. Keeping a relatively small amplitude as pure THz emitter [see Fig. \ref{Fig2}(d)], the Kerr response mainly acts by changing the pulse spectrum. More phase-matched components increase the contribution of four-wave mixing to the THz spectrum and can constructively interfere with those driven by plasma generation. In Figs. \ref{Fig2}(b,d), we refind the typical Kerr spectral signature in $\nu^2$ \cite{Borodin:ol:38:1906,Berge:prl:110:073901} that amplifies larger frequencies in the THz spectrum. We can also invoke the self-action of a THz seed originating from one potential source (e.g., Kerr or plasma) and augmenting the overall THz spectrum \cite{Cabrera:njp:17:023060}. Note the important increase in the radiated field emerging at low laser intensity in Fig. \ref{Fig2}(b). We observed that variations in the phase angle remain small and cannot solely switch on plasma emitters. So we suspect here that non-trivial couplings between the backward and forward THz components may enhance the THz yield in the low-frequency limit, which is beyond the scope of the present paper.\\

So far, linear dispersion has been discarded over propagation ranges $\leq 1$ cm along which its action is usually expected to be small in gases \cite{Berge:rpp:70:1633}. However, over comparable ranges and depending on the pump wavelength, it may already significantly impact the relative phase $\varphi$ that conditions the efficiency of the THz emitters. To illustrate this point, Figs. \ref{Fig2}(e,f) display the evolution of the phase angle $\varphi$ numerically extracted from the forward component of the electric field along the $z$ axis. The forward electric field is computed from the 1D version of the UPPE model [see Eq. (\ref{1})] for our two pulse configurations that now undergo air dispersion as modeled in Ref. \cite{Peck:josa:62:958}. At low intensity [25 TW/cm$^2$ - Fig. \ref{Fig2}(e)], for which the plasma response is small, linear dispersion drags the relative phase out of its initial value upon short distances $\sim 1.5$ cm by a phase shift comparable to that driven by the Kerr nonlinearity. A similar conclusion applies to the plasma regime [100 TW/cm$^2$ - Fig. \ref{Fig2}(f)]. Despite the smallness of its coefficients in air \cite{Peck:josa:62:958}, linear dispersion induces a long phase mismatch between the 800-nm and 400-nm pulse components, which, combined with the nonlinearities, is able to drive a phase shift close to $\pi/2$ over 1.5 cm that cannot be neglected. This constraint is relaxed to some extent for longer pump wavelengths.\\

In summary, the above results show a net influence in the THz yield when accounting for the Kerr nonlinearity along cm-propagation ranges. The Kerr response directly alters the pump spectrum as well as variations in the phase angle between the $\omega-2\omega$ pulse components. Together with linear dispersion, this impacts the THz conversion efficiency, which is mainly driven by the photocurrent mechanism at clamping intensity.\\

\section{Impact of delayed Kerr nonlinearities and longer pump wavelength} \label{S3}

Below we address the influence of the Raman-delayed Kerr nonlinearity on the laser-to-THz conversion efficiency and we quantify the increase in THz generation when the fundamental wavelength belongs to the mid-infrared range. Since from Fig. \ref{Fig1} we expect no significant action from the longitudinal field $E_z$, we only focus on the transverse field $E_x$, whose subscript $x$ is henceforth omitted.

\subsection{Influence of the Raman-delayed nonlinearity}

For notational convenience, we rewrite the initial laser field (\ref{e2}) as
\begin{equation}
\label{eq:ea}
E(t) =  \mathcal{E}_{\omega_0}(t) a_{\omega_0} \cos(\omega_0 t) + \mathcal{E}_{2\omega_0}(t) a_{2\omega_0} \cos(2\omega_0 t + \varphi),
\end{equation}
where $0 \leq \mathcal{E}_{\omega_0,2\omega_0}(t) \leq 1$, $a_{\omega_0,2\omega_0}$ and $\varphi$ are the pulse envelopes with
duration $\tau_{\omega_0,2\omega_0}$, relative amplitude at $\omega_0$ or $2\omega_0$ and the relative phase between the two colors, respectively. We assume long enough FWHM durations, i.e., $\omega_0 \tau_{\omega_0,2\omega_0} \gg 1$.\\

Using the input two-color pulse (\ref{eq:ea}), we can evaluate the low-frequency part of the overall Kerr response Eq. (\ref{kerrdel}), when both envelope functions $\mathcal{E}_{\omega_0,2\omega_0}(t)$ take the value unity for the sake of simplicity. Cook and Hochstrasser's result \cite{Cook:ol:25:1210} is easily recovered for the instantaneous part of the Kerr polarization yielding the direct-current (dc) contribution
\begin{equation}
\label{kerrTHz}
P_{\rm inst}^{\rm dc} = (1-x_k) \frac{3}{4} \epsilon_0 \chi^{(3)} a_{\omega_0}^2 a_{2\omega_0} \cos{\varphi}.
\end{equation}
THz emission due to an instantaneous Kerr response is maximum for the phase offset $\varphi = 0$ $[\pi]$.\\

Adding the Raman contribution now leads us to evaluate the integral
\begin{equation}
\label{Raman1}
P_{\rm Raman} = x_k \frac{3}{2} \epsilon_0 \chi^{(3)} E(t) \int_0^{+\infty} \frac{\tau_1^2+\tau_2^2}{\tau_1 \tau_2^2} \mbox{e}^{-\tau/\tau_2} \sin{\left( \frac{\tau}{\tau_1} \right)}E^2(t-\tau) d\tau.
\end{equation}

\begin{figure}
\centering \includegraphics[width=\columnwidth]{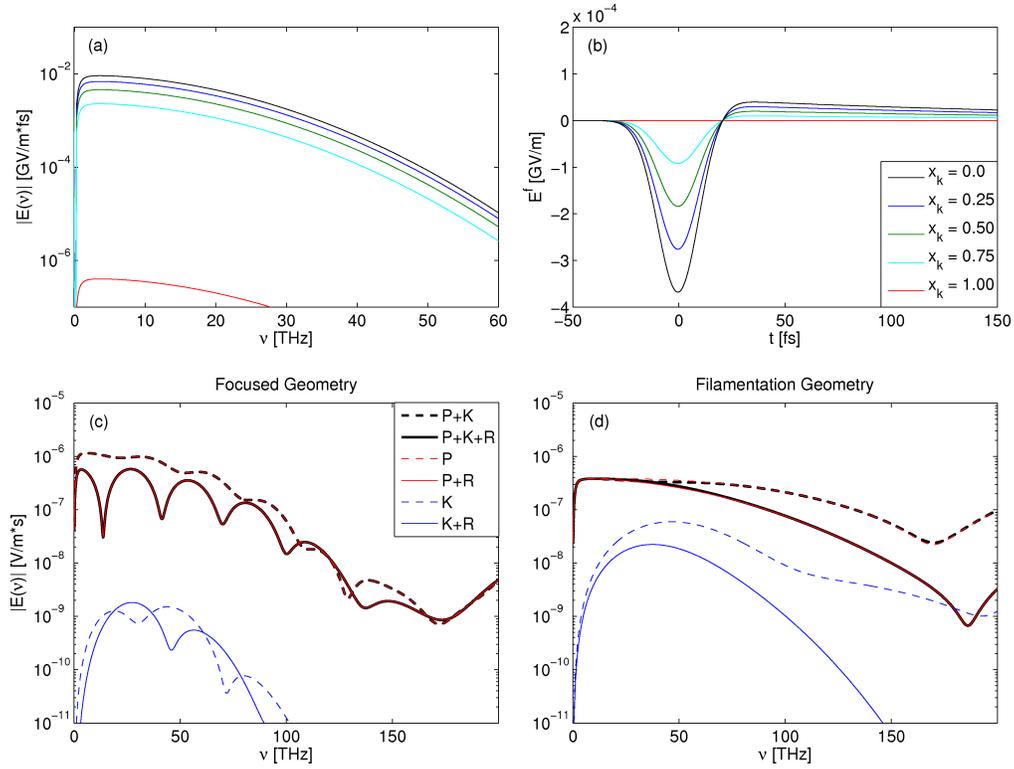}
\caption{(a) THz spectra as functions of the fraction $x_k$ of Raman nonlinearity with no ionization. (b) Corresponding fields emitted by four-wave mixing and filtered in a 80-THz-wide window. (c,d) THz spectra computed by plugging the temporal profiles of two-color pulses into the source terms $\partial_t J$ and/or $\partial_t^2 P_{\rm NL}$, isolated or summed up with (R) or without the Raman contribution (see legend) for the pulse configurations using a 800-nm pump simulated in (c) Fig. \ref{Fig7a} at the distance $z = 2.5$ cm and (d) Fig. \ref{Fig8}(a) at the distance $z = 20$ cm.}
\label{Fig3}
\end{figure}

After several trigonometric simplifications, we can extract from the previous expression a low-frequency (dc) contribution reading as
\begin{equation}
\label{Raman2}
P_{\rm Raman}^{\rm dc} = x_k \frac{3}{2} \epsilon_0 \chi^{(3)} a_{\omega_0}^2 a_{2\omega_0} [ T_1 \cos{\varphi} + T_2 \sin{\varphi}],
\end{equation}
where 
\begin{equation}
\label{T1}
T_1 = \frac{\tau_1^2 + \tau_2^2}{4} \left(\frac{\tau_2^2 + \tau_1^2(1 - 4 \tau_2^2 \omega_0^2)}{\alpha(\tau_1,\tau_2,\omega_0)} + \frac{2 \tau_1 \omega_0(\tau_2^2 - \tau_1^2(1 + \tau_2^2 \omega_0^2))}{\beta(\tau_1,\tau_2,\omega_0)}\right),
\end{equation}
\begin{equation}
\label{T2}
T_2 = \frac{\tau_1^2 + \tau_2^2}{4} \left(- \frac{4 \tau_1^2 \tau_2 \omega_0}{\alpha(\tau_1,\tau_2,\omega_0)} + \frac{2 \tau_1 \tau_2^{-1}(\tau_2^2 + \tau_1^2(1 + \tau_2^2 \omega_0^2))}{\beta(\tau_1,\tau_2,\omega_0)}\right),
\end{equation}
with
\begin{equation}
\label{alpha}
\alpha(\tau_1,\tau_2,\omega_0) = (\tau_1^2 + \tau_2^2)^2 + (2\tau_1 \tau_2 \omega_0)^2(2 \tau_1^2 - 2 \tau_2^2 + 4\tau_1^2 \tau_2^2 \omega_0^2),
\end{equation}
\begin{equation}
\label{beta}
\beta(\tau_1,\tau_2,\omega_0) = (\tau_1^2 + \tau_2^2)^2 + (\tau_1 \tau_2 \omega_0)^2(2 \tau_1^2 - 2 \tau_2^2 + \tau_1^2 \tau_2^2 \omega_0^2).
\end{equation}
The values of $T_1$ and $T_2$ provide an optimal phase for THz generation when $\varphi = \tan^{-1} (T_2/T_1)$. For pump wavelengths $\lambda_0$ comprised between 0.8 and 3 $\mu$m, we find that $|T_1| \leq 0.021$ and $|T_2| \sim 0.02 |T_1|$. It is thus reasonable to neglect $T_2$, so that the low-frequency part of the total Kerr response involving the Raman-delayed component simplifies into
\begin{equation}
\label{TotalKerrTHz}
P_{\rm NL}^{\rm dc} \approx \frac{3}{4} \epsilon_0 \chi^{(3)} a_{\omega_0}^2 a_{2\omega_0} [1 - x_k(1-2 T_1)] \cos{\varphi}.
\end{equation}
By analogy with Cook and Hochstrasser's analysis, this formula indicates that the optimal phase difference for THz generation by the Kerr response remains $\varphi=0$ $[\pi]$. Maximum THz generation is provided when there is no Raman nonlinearity $(x_k = 0)$. $P_{\rm Raman}$ decreases the THz yield by a correction $(T_1)$ of the percent order, which is confirmed by Figs. \ref{Fig3}(a,b) for different values of the fraction $x_k$. THz spectra and fields are computed at $z=0$ from Eq. (\ref{e6}) including the nonlinear polarization. They indeed decrease in amplitude when $x_k$ is augmented. This property reflects the fact that the nonlinear integrand in Eq. (\ref{Raman1}) acts over relatively long relaxation times $\tau_1-\tau_2 \sim 60-80$ fs, along which the high-frequency laser oscillations cancel each other over the integration in time. The resulting integral is slowly varying and $P_{\rm Raman}$ thus barely contributes to the THz spectrum.\\

This result confirms the behavior expected from envelope-like unidirectional models \cite{Berge:rpp:70:1633} for which the Raman nonlinearity is assumed insensitive to the pump harmonics and oscillates likely to the optical field. It justifies that we can employ Eq. (\ref{kerrdel}) within a field description and still applies to alternative formulations of the rotational Raman scattering \cite{Palastro:pra:86:033834}. Besides the net decrease by the fraction $x_k$, the impact of the overall Kerr source is also expected to decrease from the loss of self- and cross-phase modulations that affect the pump spectrum through the Kerr response. This aspect is detailed in Figs. \ref{Fig3}(c,d), where we have compared the THz spectra obtained from inserting into the source terms $\partial_t J$ and/or $\partial_t^2 P_{\rm NL}$ some temporal profiles further computed in Section \ref{S4} in focused geometry [Fig. \ref{Fig3}(c)] and filamentation regime [Fig. \ref{Fig3}(d)] at the distance of maximum THz generation. Although this procedure loses the memory of the THz yield accumulated along previous distances, including that prior to ionization, it clearly indicates that in plasma regime the Kerr source remains minor compared to that associated to photocurrents. The action of the Raman nonlinearity mostly manifests by changing the pulse spectrum, which conditions the photocurrents.\\

\subsection{Increase in the pump wavelength}

References \cite{Berge:prl:110:073901,Clerici:prl:110:253901} reported an impressive increase in the THz conversion efficiency when doubling the fundamental wavelength for equal FWHM durations. For single-color pulses of same energy, we can easily expect that doubling $\lambda_0$ for THz generation triggered in tunneling regime keeps the final electron density $N_e$ unchanged, but it doubles the free electron velocity \cite{Babushkin:njp:13:123029}
\begin{equation}
\label{eq:vf}
v_f(t) = -\frac{e}{m_e}{\rm e}^{-\frac{t}{\tau_c}}\int_{-\infty}^tE(t'){\rm e}^{\frac{t'}{\tau_c}}\,dt'
\end{equation}
through its evident dependency on $1/\omega_0$. Thereby the electron current density is doubled. With two colors involving a dominant fundamental pulse at $\omega_0$, the same scaling holds. However, $N_e$ noticeably increases, e.g., by a factor $\sim 2$ at 100 TW/cm$^2$, for two superimposed colors ($\varphi=0$). With a $\pi/2$ relative phase, the pump field exhibits a temporal asymmetry around the electric field maxima and the current density $J(t)$ develops a low-frequency component due to the stepwise increase of the electron density. This component is then the major THz source \cite{Li:apl:101:161104}.\\

Following the local current theory \cite{Babushkin:njp:13:123029}, ionization happens near the relative extrema of $E(t)$ at the instants $t_1,t_2,t_3,\ldots t_n$, from which the electron density and current can be approximated as $N_e(t) \simeq \sum_n\delta N_n H_n(t-t_n)$ and $J(t) \simeq J_A(t) + J_B(t)$ with
\begin{equation}
\label{eq:JA}
J_A(t) =  -e \sum_n \delta N_n  v_f(t) H_n(t-t_n),\,\,\,
J_B(t) = e \sum_n \delta N_n {\rm e}^{-\frac{t-t_n}{\tau_c}}v_f(t_n) H_n(t-t_n).
\end{equation}
The quasi-step function is $H_n(t)=\frac{1}{2}[1+ \mbox{erf}(t/\tau_n)]$, where the duration of the $n$th ionization event is $\tau_n = \left[3(U_H/U_i)^{3/2} |E(t_n)|^2/(|\partial_t^2{E}(t_n)|E_a) \right]^{1/2}$. Using Fourier transforms, we obtain in the low-frequency domain and in the non-collisional limit (see Refs. \cite{Babushkin:njp:13:123029,Gonzalez:prl:114:183901}):
\begin{equation}
\label{eq:freqJA}
\mathcal{F}[\partial_t J_A](\omega) \approx \frac{- {\rm i} e^2}{\sqrt{2\pi} m_e} \sum_n \delta N_n \left( E_{\omega_0}(t_n) + \frac{1}{4} E_{2\omega_0}(t_n) \right) \frac{\omega}{\omega_0^2},
\end{equation}
\begin{equation}
\label{eq:freqJB}
\mathcal{F}[\partial_t J_B](\omega) \approx \frac{e}{\sqrt{2\pi}}\sum_n\delta N_n v_f(t_n){\rm e}^{{\rm i} \omega t_n}, 
\end{equation}
where $E_{\omega_0}(t)=\mathcal{E}_{\omega_0}(t) a_{\omega_0} \cos(\omega_0 t)$ and $E_{2\omega_0}(t)=\mathcal{E}_{2\omega_0}(t) a_{2\omega_0} \cos(2\omega_0 t + \varphi)$.\\

Ignoring again the influence of the envelopes $(\mathcal{E}_{\omega_0,2\omega_0}=1)$, the ionization instants of a two-color pulse can be approximated by \cite{Babushkin:njp:13:123029}
\begin{equation}
\label{tn}
\omega_0 t_n \approx n \pi - 2 \frac{a_{2\omega_0}}{a_{\omega_0}} (-1)^n \sin{\varphi}
\end{equation}
at leading order in $a_{2\omega_0}/a_{\omega_0} \ll 1$. We assume equal density jumps
\begin{equation}
\label{eq:rhon}
\delta N_n \simeq N_a \left(1-{\rm e}^{-\sqrt{\pi}W[E(t_1)]\tau_1} \right)\equiv \delta N,
\end{equation}
occurring over a large number of ionization events $N \gg 1$. In the THz frequency range $\omega \ll \omega_0$, it is then straightforward to evaluate
\begin{equation}
\label{TFA}
\mathcal{F}[\partial_t J_A](\omega) \approx - \frac{{\rm i} e^2 a_{2\omega_0}}{4\sqrt{2\pi} m_e} N \delta N \frac{\omega}{\omega_0^2} \cos{\varphi},
\end{equation}
\begin{equation} 
\label{TFB}
\mathcal{F}[\partial_t J_B](\omega) \approx \frac{3 e^2 a_{2\omega_0}}{\sqrt{2\pi} m_e} \frac{\delta N}{\omega_0} \frac{\sin{(N \omega \pi/2 \omega_0)}}{\sin{(\omega \pi/\omega_0)}}  \sin{\varphi},
\end{equation}
\begin{equation}
\label{TF}
\mathcal{F}[\partial_t J](\omega) \approx \frac{e^2}{\sqrt{2\pi} m_e} a_{2\omega_0} \frac{\delta N}{\omega_0} \left[ - \frac{{\rm i}}{4} N \frac{\omega}{\omega_0} \cos{\varphi} + 3 \frac{\sin{(N \omega \pi/2 \omega_0)}}{\sin{(\omega \pi/\omega_0)}}  \sin{\varphi} \right].
\end{equation}
Equations (\ref{TFA}) and (\ref{TFB}) show that $\partial_t J_A$ and $\partial_t J_B$ dominate for $\varphi=0$ and $\varphi=\pi/2$, respectively. The ionization steps $\delta N$ [Eq. (\ref{eq:rhon})] increase, in the limit $W \tau_n \ll 1$, linearly with the ionization duration $\tau_n \approx \tau_1$. For a fixed pulse duration, the number of optical cycles is halved when one doubles the pump wavelength and since $\tau_n \propto \omega_0^{-1}$, one has $\tau_{2\lambda_0}/\tau_{\lambda_0} \sim 2$. The number of ionization events decreases accordingly, i.e., $N_{2\lambda_0}/N_{\lambda_0} = 1/2$.\\

For a better understanding of $\partial_t J_A$, it may be instructive to rewrite Eq. (\ref{eq:freqJA}) in the form
\begin{equation}
\label{eq:freqJAnew}
\mathcal{F}[\partial_t J_A](\omega) \approx \frac{- {\rm i}\omega e}{\sqrt{2\pi}} \sum_n \delta N_n r_f(t_n),
\end{equation}
where $r_f$ denotes the free electron position ($r_f=0$ at $-\infty$):
\begin{equation}
\label{eq:rf}
r_f(t) = \int_{-\infty}^tv_f(t')\,dt'.
\end{equation}

\begin{figure}[h]
\centering \includegraphics[width=\columnwidth]{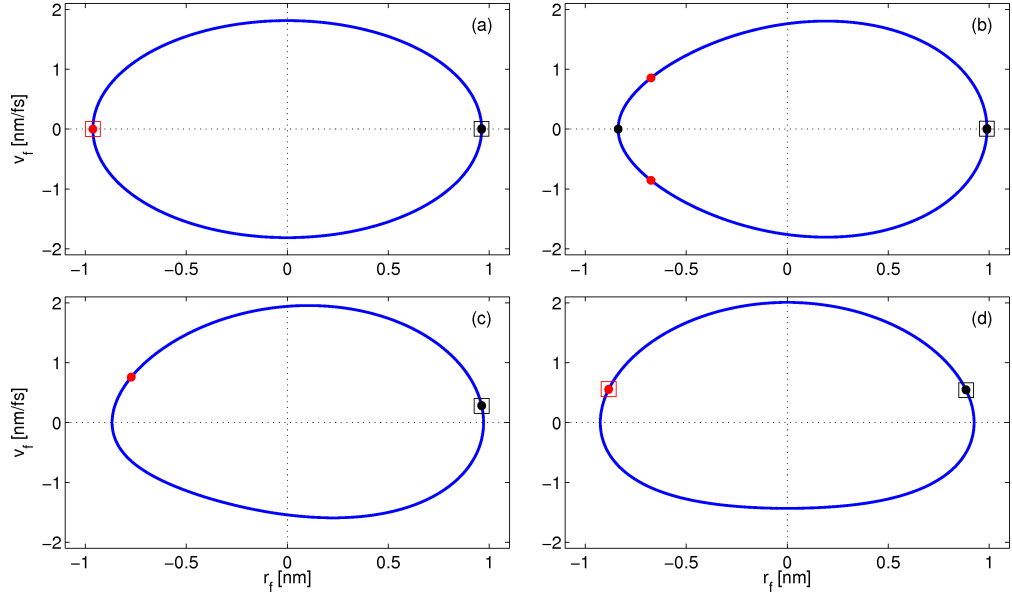}
\caption{Velocity of free electrons created from $t=-\infty$ as function of their position. The dots locate the minima (red dots) and maxima (black dots) of the laser field. (a) Single color. (b) Two colors with zero relative phase. (c) Two colors with a relative phase of $\pi/4$. (d) Two colors with a relative phase of $\pi/2$. Squares indicate the location of the strongest extrema for which $\delta N_n$ exceeds the others by more than one order of magnitude.}
\label{Fig4}
\end{figure}

Figure \ref{Fig4} illustrates the phase space $(r_f,v_f)$. It provides a qualitative way to rapidly know if a given pulse configuration favors THz generation from $J_A$ or $J_B$. Dots correspond to the minima and maxima of the laser field at the coordinates $[r_f(t_n),v_f(t_n)]$. There are constructive contributions if $r_f(t_n)$ for $\mathcal{F}[\partial_t J_A]$ or $v_f(t_n)$ for $\mathcal{F}[\partial_t J_B]$ are sign-definite. For one color [Fig. \ref{Fig4}(a)], there is only one contribution of $\mathcal{F}[\partial_t J_A]$ which is destructive: $v_f(t_n)$ is zero, while the symmetric extrema in the positions $r_f(t_n)$ cancel each other due to their opposite signs. For two colors with a null relative phase [Fig. \ref{Fig4}(b)], again destructive contributions exist in $\mathcal{F}[\partial_t J_A]$ and $\mathcal{F}[\partial_t J_B]$, but the configuration is more favorable to $\mathcal{F}[\partial_t J_A]$ for which $r_f(t_n)$ cannot cancel out. For two colors and $\varphi=\pi/2$ [Fig. \ref{Fig4}(d)], positive velocities $v_f(t_n)>0$ enhance $\mathcal{F}[\partial_t J_B]$, whereas $\mathcal{F}[\partial_t J_A]$ vanishes with opposite $r_f(t_n)$. This situation also applies to Fig.~\ref{Fig4}(c).\\

\begin{figure}[h]
\centering \includegraphics[width=\columnwidth]{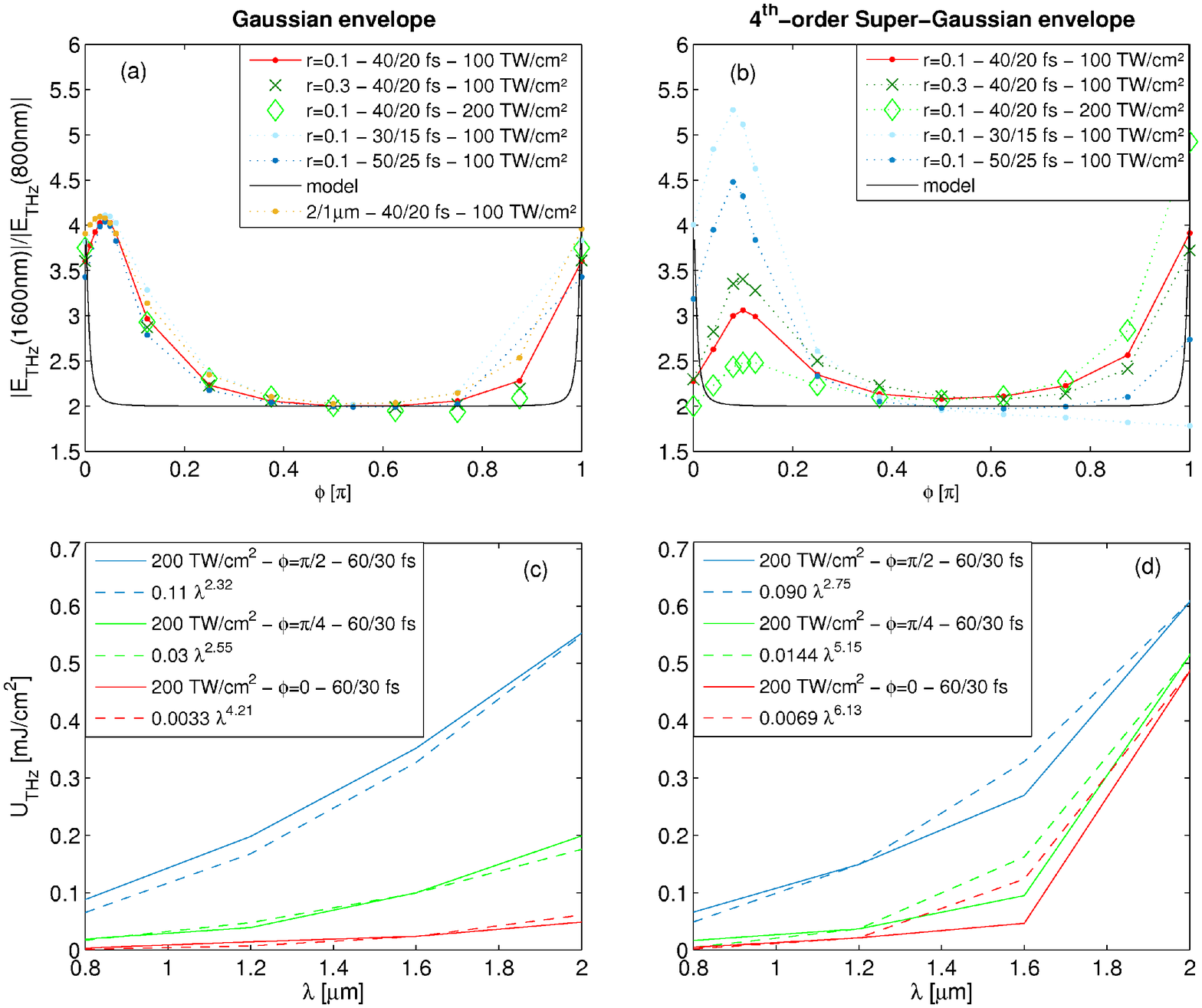}
\caption{(a,b) Ratio of THz field strengths emitted by a two-color pulse with a 1600-nm pump over that emitted with a 800-nm pump, function of the relative phase between the fundamental and second harmonic in a 80-THz-wide window. (a) Gain factor for Gaussian pulses with various intensity ratios $r \simeq a_{2\omega_0}^2/a_{\omega_0}^2$ and main laser intensities $I_0$ (see legend). Blue dotted lines show the gain factor for two different pulse durations at the same intensity. The black solid line shows the gain factor using the analytical model Eq. (\ref{TF}). (b) Same quantity for 4th-order super-Gaussian envelopes with different ratios $r$ and intensities $I_0$. (c,d) Evolution of the THz yield with the pump wavelength for the three relative phases 0, $\pi/4$ and $\pi/2$ using (c) Gaussian pulses and (d) 4th-order super-Gaussian pulses.}
\label{Fig5}
\end{figure}

In Figs. \ref{Fig4}(a,b), stricto-sensu, THz emission should not be zero due to the pulse envelope \cite{Babushkin:njp:13:123029}. Considering the influence of bounded envelopes, Fig. \ref{Fig5} shows the ratio between the THz field induced by a two-color pump-pulse with fundamental wavelength at 1600 nm and one with fundamental wavelength at 800 nm. This ratio is evaluated through an ordinary least squares method applied to the THz field profiles calculated numerically from the LC model for ionizing beam intensities. This method mostly captures the ratio between the THz field maxima along the time axis. For comparison, the theoretical ratio $|{\tilde E}_{2\lambda_0}/{\tilde E}_{\lambda_0}|$ inferred from the inverse Fourier transform of Eq. (\ref{TF}) is plotted as a black solid line. It provides a gain factor that varies between 4 ($\varphi=0$) and 2 ($\varphi=\pi/2$) with $\pi$-periodicity. For Gaussian pulses [$\mathcal{E}_{\omega_0,2\omega_0}(t)=\exp{(-2^{2\beta-1}\ln{2}\,t^{2\beta}/\tau_{\omega_0,2\omega_0}^{2\beta})}$ with $\beta = 1$], this behavior is verified by the LC results (red curves) of Fig. \ref{Fig5}(a), despite minor variations caused by envelope effects. The ratio $|{\tilde E}_{2\lambda_0}/{\tilde E}_{\lambda_0}|$ remains less sensitive to the pulse duration and the fundamental wavelength than to the relative phase $\varphi$. Maximum gain factor in the THz field amplitudes is obtained for $\varphi \approx 0$, which underlines the role of the current density $J_A$ directly connected to the laser field. Opting next for 4th-order super-Gaussian profiles $(\beta = 4$), $|{\tilde E}_{2\lambda_0}/{\tilde E}_{\lambda_0}|$ again varies with the relative phase, but it strongly evolves and even exceeds the value~5 for $\varphi \leq \pi/10$ when decreasing the FWHM duration [Fig. \ref{Fig5}(b)]. We attribute these changes to the steepness of the envelope, which makes the first ionization events not exactly located at the same times for 800-nm and 1600-nm pump pulses. For our laser parameters, the relative intensity ratio $r$ appears to have a limited impact on the gain performances.\\

Figures \ref{Fig5}(c,d) illustrate the increase in the THz energy yield for 60-fs FWHM pump duration and intensities $\sim 200$~TW/cm$^2$ rather reached in focusing geometries \cite{Clerici:prl:110:253901}. Solid lines refer to the computed energy values, while the dashed lines are fitting curves in $\lambda^{\alpha}$. We can observe that exponents $\alpha>4$ fit for small relative phases. A $\pi/2$ phase angle, however, renders the $J_B$ contribution dominant and thus decreases this exponent. So, even though $\lambda$-dependent scalings reported in \cite{Clerici:prl:110:253901} are possible, they are not generic as the gain factors are highly sensitive to the relative phase between the two colors, the shape of the pulse envelopes and their durations. Note from Figs. \ref{Fig5}(c,d) that the THz pulse energy is much larger with a $\pi/2$ phase angle than with a null phase. This means that in the situation where $J_A$ is dominant $(\phi \rightarrow 0)$, the overall THz spectrum is much smaller than when $J_B$ prevails for different phase values.\\

\section{Comparison with unidirectional pulse propagation simulations} \label{S4}

The previous properties are now checked by direct 3D numerical computations. 
Our reference model is the unidirectional pulse propagation equation (UPPE)~\cite{Kolesik:prl:89:283902,Kolesik:pre:70:036604} that governs the forward-propagating component of linearly polarized pulses
\begin{equation}
 \partial _{z}\hat{E}= {\rm i} \sqrt{k^{2}(\omega)-k_{x}^{2}-k_{y}^{2}}\, \hat{E} + {\rm i} \frac{\mu _{0}\omega ^{2}}{2k(\omega )}\hat{\mathcal{F}}_{\mathrm{NL}},
\label{1}
\end{equation}
where $\hat{E}(k_{x},k_{y},z,\omega )$ is the Fourier transform of the laser electric field with respect to $x$,
$y$, and $t$. The first term on the right-hand side of Eq.~(\ref{1})
describes linear dispersion and diffraction of the pulse. The
term $\hat{\mathcal{F}}_{\mathrm{NL}}=\hat{P}_{\rm NL}+ {\rm i} \hat{J}/\omega + {\rm i} \hat{J}_{\mathrm{loss}}/\omega$ contains the third-order nonlinear polarization with Kerr index
$n_{2}=3\chi^{(3)}/4n_0^2c\epsilon_0$ $[n_0 = n(\omega_0)]$, the electron current ${J}$ and a loss
term $J_{\mathrm{loss}}$ due to ionization \cite{Berge:rpp:70:1633,Berge:prl:100:113902}. 
Compared with Ref. \cite{Kolesik:pre:70:036604}, the denominator of the nonlinear term reduces to $2 k(\omega)$, assuming $\omega^2\hat{\mathcal{F}}_{NL}$  relevant only for $k(\omega) = n(\omega) \omega/c \gg \sqrt{k_x^2+k_y^2}$.\\

The coming analysis aims at first validating our theoretical expectations using the simple QST model for a single species (O$_2$) and classical values for the Kerr indices, in both focused and parallel propagation geometries. Next, more elaborated ionization models and recently measured Kerr coefficients in the mid-infrared will be employed to countercheck our findings.\\

\subsection{Validation of theoretical issues}

Equation (\ref{1}) is here solved for experimental configurations close to those examined in \cite{Clerici:prl:110:253901}, i.e., for focused pulses with $f$-numbers $> 10$ ($f$-number refers to the ratio between the focal length and the FWHM input beam diameter). The initial relative phase $\varphi$ is set equal to zero and the fundamental pump wavelengths of the two-color pulses are 800 nm and 1600 nm.\\

In a first set of simulations we choose $f/\# = 42$ for the beam width $w_0 = 500$ $\mu$m and focal length $f=2.5$ cm. Our two-color pulses have 200 $\mu$J in energy and the FWHM duration of the pump pulse is 60 fs. About $7\,\%$ of the laser energy is contained in the second harmonic. Dispersion curve in air for the refractive optical index $n(\omega)$ is again taken from Ref.~\cite{Peck:josa:62:958}. Kerr indices are chosen as $n_2 \simeq 1.2 \times 10^{-19}$ cm$^2$/W following Refs.~\cite{Loriot:oe:17:13429,Ettoumi:oe:18:6613}. At atmospheric pressure, $N_a = 5.4 \times 10^{18}$ cm$^{-3}$ for O$_2$ molecules and the critical power for self-focusing, defined by $P_{\rm cr} \simeq\lambda_0^2/2\pi n_0n_2$, is $P_{\rm cr} = 8.5$ GW at 800 nm and 35.1 GW at 1600 nm. Although strongly focused, our ultrashort pulses promote single-ionization events for peak intensities $< 300$ TW/cm$^2$ near focus due to local defocusing by the generated plasma. Therefore, we can here use the QST ionization rate (\ref{QST}) applied to oxygen molecules only.\\

\begin{figure}[h]
\centering \includegraphics[width=\columnwidth]{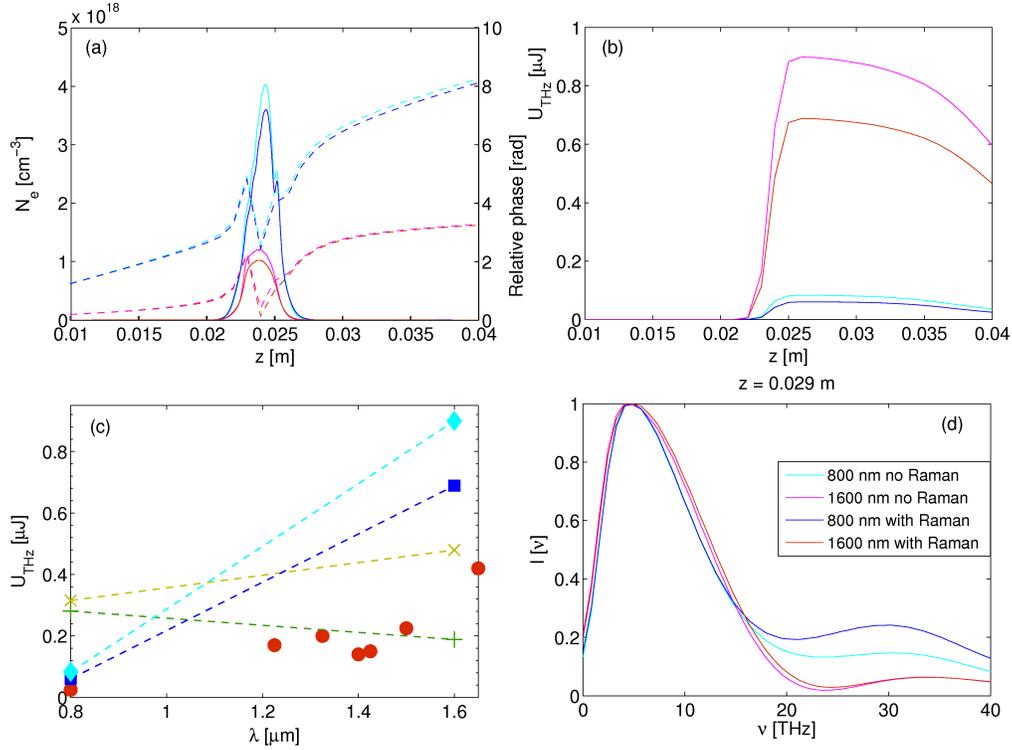}
\caption{Focused two-color Gaussian pulses in ratio $r = 7.4\%$. (a) Maximum electron density (left-hand side axis, solid curves) and relative phase between the two pulse components (right-hand side axis, dashed curves), (b) THz energy yield ($\nu \leq 80$ THz) for the pump wavelengths 800 nm (blue/cyan curves) and 1600 nm (red/magenta curves), with and without the Raman nonlinearity. (c) THz energy vs pump wavelength. Cyan diamonds: no Raman; blue squares: With Raman. Green crosses $\times$ report THz gain factors in filamentation regime with no Raman nonlinearity as promoted in Fig. \ref{Fig8}; Green symbols $+$ report gain factors in filamentation regime with Raman nonlinearity as given by Fig. \ref{Fig14}. The solid red circles recall the experimental data points of Ref. \cite{Clerici:prl:110:253901}. (d) Normalized THz spectra computed near focus.}
\label{Fig7a}
\end{figure}

Simulations have been performed with a time window of 1.22 ps, a temporal step $\Delta t = 75$~attoseconds and transverse resolution of $\Delta x = \Delta y \approx 3$ $\mu$m. Figure \ref{Fig7a} shows the peak electron density reached near focus, the variations of the relative phase between the fundamental and second harmonic along $z$ [Fig. \ref{Fig7a}(a)], and the THz energy contained in our numerical box ($3 \times 3$ mm$^2$) [Fig. \ref{Fig7a}(b)]. THz radiation is collected within a 80-THz-large frequency window. Cyan/magenta curves ignore the delayed Raman nonlinearity; blue/red curves include it for comparison $(x_k=0.5)$. There is a limited influence of the Kerr response in tight focusing regime, but its related THz conversion efficiency is clearly diminished by the delayed nonlinearity for the reasons given in Section \ref{S3}. Concerning the wavelength dependency, the maximum intensity achieved near focus decreases with $\lambda_0$, as the beam waist $w_f \approx w_0 f/z_0$ becomes proportional to the pump wavelength when the Rayleigh length $z_0 = \pi w_0^2/\lambda_0$ is much larger than the focal distance $f$ \cite{Berge:pra:88:023816}. Consequently, since the QST rate (\ref{QST}) does not depend on the pump wavelength, the peak electron density decreases in turn [Fig. \ref{Fig7a}(a)]. The relative phase $\varphi$ covers the full interval $[0,2\pi]$ over the 4-cm-long propagation range. It experiences a Gouy phase shift up to $\pi$ near focus, supplemented by another $\pi$ phase shift induced by linear dispersion for the 800-nm pump [see Fig. \ref{Fig2}(f)]. In Fig. \ref{Fig7a}(b) we observe a net increase of the maximum THz energy produced at $z \simeq f$ when $\lambda_0$ is augmented.\\

The experimental data points of Ref. \cite{Clerici:prl:110:253901} are recalled by solid red circles in Fig. \ref{Fig7a}(c), which we compare with the THz pulse energy at focus. Despite differences between the original experiment and our laser parameters, the THz yield only evaluated from two pump wavelengths follows a comparable growth. Accounting for Raman scattering helps to reach a better agreement with the experimental data. For our THz window of 80 THz, a rough fitting curve indicates a growth rate in $\lambda^{\alpha}$ with $\alpha \approx 3.5$, i.e., $2 < \alpha \leq 4$ in agreement with Fig. \ref{Fig5}(c,d), keeping in mind the variations in the relative phase $\varphi$ shown in Fig. \ref{Fig7a}(b). Shortening this window to 20 THz as measured in \cite{Clerici:prl:110:253901} does not noticeably change this scaling, as the THz spectra emitted around focus are self-contained in the frequency domain $\nu <20$ THz [Fig. \ref{Fig7a}(d)]. The spectral distributions remarkably support the comparison with the measurements of \cite{Clerici:prl:110:253901}, in particular by achieving a maximum at $\nu = 5$ THz. The influence of the plasma volume is here limited: Visual inspection of the numerical data indeed revealed plasma channels being of comparable dimensions when using a pump wavelength of either 800 nm or 1600 nm, i.e., the plasma volumes only vary within a factor 0.8 - 1.3 from density levels $> 10^{15-17}$ cm$^{-3}$.\\

We now employ Eq. (\ref{1}) to describe THz generation by two-color filaments operating with two different pump wavelengths over long distances. The UPPE is integrated for two-color Gaussian pulses with input power $P_{\rm in}=34$ GW, beam waist $w_0=400$ $\mu$m and FWHM durations $\tau_{\omega_0} = 40$ fs ($\tau_{2\omega_0} = \tau_{\omega_0}/2$, $r=3.4\%$) in collimated propagation. Kerr indices are unchanged and we keep the quasi-static tunneling rate (\ref{QST}). For reasons of computational cost, the numerical resolution has been decreased to $\Delta t = 99$ attoseconds and $\Delta x = \Delta y \approx 9$ $\mu$m, which was checked to introduce no significant variations in the THz spectra and fields. The selected THz window is still $\nu \leq 80$ THz.\\

\begin{figure}
\centering \includegraphics[width=\columnwidth]{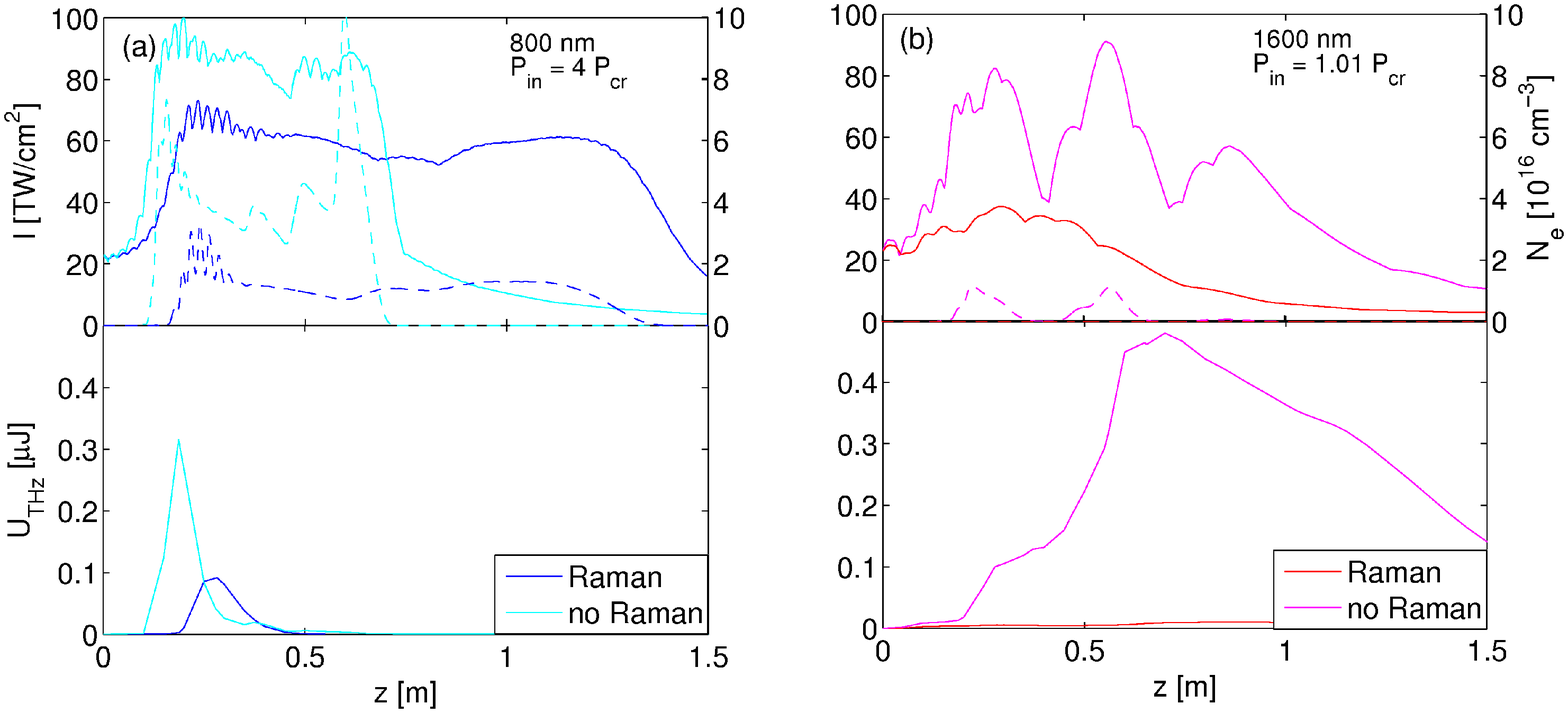}
\centering \includegraphics[width=\columnwidth]{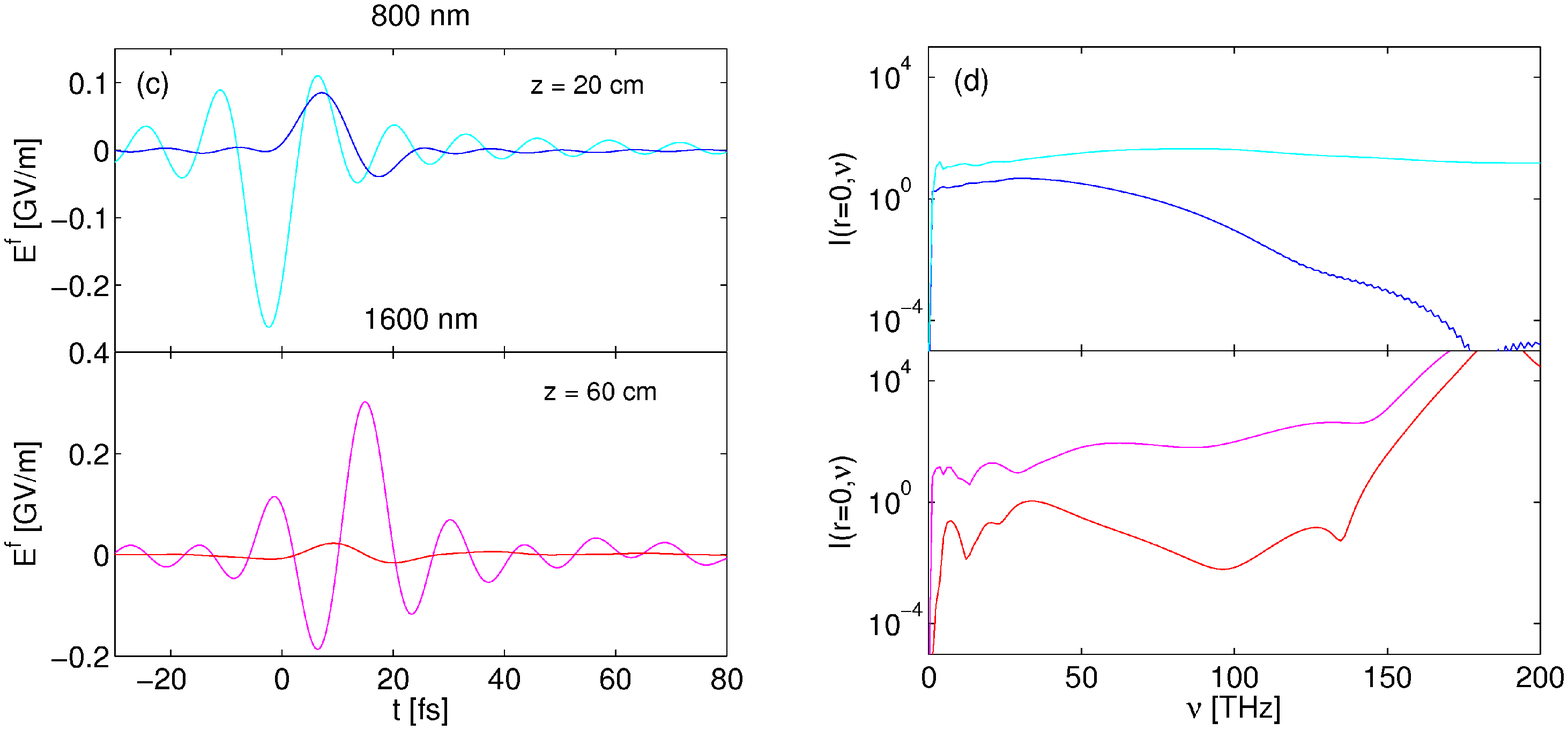}
\caption{(a) Top: Maximum intensity (solid curves, left-hand side axis) and peak electron density (dashed curves, right-hand side axis) of a meter-long two-color filament with 800-nm (blue/cyan curves) and 1600-nm pump component (red/magenta curves) $(r=3.4\%)$. Colored curves refer to wave propagation without (cyan/magenta curves) and with (blue/red curves) Raman-delayed Kerr nonlinearity. Bottom: Corresponding THz pulse energy yield accumulated along $z$ inside a 80-THz-large frequency window. (c,d) On-axis THz fields and intensity spectra near the distance of maximum THz energy yield.}
\label{Fig8}
\end{figure}

Figures \ref{Fig8}(a,b) illustrate the peak intensity and maximum electron density reached along meter-range distances in two-color filamentation regime. Light (cyan and magenta) curves show a propagation for which no delayed Kerr nonlinearity is accounted for ($x_k=0$). Dark (blue and red) curves include the Raman-delayed nonlinearity in ratio $x_k=0.5$ \cite{Pitts:josab:21:2008,Champeaux:pre:71:046604,Champeaux:pre:77:036406}. From Fig. \ref{Fig8}(a) it is clear that the Raman term weakens the contribution of the instantaneous Kerr response, which results in (i) a longer self-focusing distance, (ii) a decreased clamping intensity, and thereby (iii) an enhanced self-guiding range. Consistently, the peak plasma density decreases in turn and extends over longer distances. As expected, the bottom row of Figs. \ref{Fig8}(a,b) displays a net decrease of the THz energy emitted along propagation. With a pump wavelength of 1600 nm, at equal energy content, the input power becomes closer to a single critical power in air and self-channeling with no delayed-Kerr response favors a more extended filamentation range. Peak densities decrease from $10^{17}$ cm$^{-3}$ to $10^{16}$ cm$^{-3}$, which should weaken the high gain factors achieved in focused geometry. Indeed, the THz yield is only stronger (without Raman), locally by a factor $\sim 1.52$ at maximum emission, as indicated by the green crosses of Fig. \ref{Fig7a}(c). Note that the THz energy with a 800-nm pump wave can escape early our numerical box ($2.4 \times 2.4$ mm$^2$), so the previous evaluation is performed at the maximum of THz energy. Performances in the THz gain factor with an increased pump wavelength in collimated propagation thus appear weaker than in focused geometry. When adding the Raman-delayed nonlinearity, the available power contributing to the instantaneous Kerr response is subcritical $(\sim 0.7)$, which prevents the two-color pulse from self-focusing and exceeding the ionization threshold. As a result, no plasma generation takes place and only a residual THz emission occurs due to four-wave mixing.\\ 

Figure \ref{Fig8}(c,d) depicts THz fields propagated over several tens of cm and their spectra for 800-nm and 1600-nm pump pulses. The selected distances correspond to the range of maximum THz energy shown in Figs. \ref{Fig8}(a,d), bottom. One clearly sees that the Raman-delayed response contributes locally to diminish the THz conversion efficiency, because the pump dynamics is changed and in particular the peak intensity is reduced. In addition, a longer pump wavelength promotes the formation of a supercontinuum linking the tail of the fundamental pulse spectrum to the THz spectrum [Fig. \ref{Fig8}(d)], which can justify a stronger influence of the current component $J_A$. THz fields with $\sim 0.1$ GV/m amplitudes are achieved at both pump wavelengths with a clear amplification at $z = 60$ cm from the 1600-nm pump.\\

\subsection{Generalization for different medium parameters}

We now test our previous findings for more complex ionization rates applying to the two major air species and consider different bound electron responses.\\

\begin{figure}
\centering \includegraphics[width=\columnwidth]{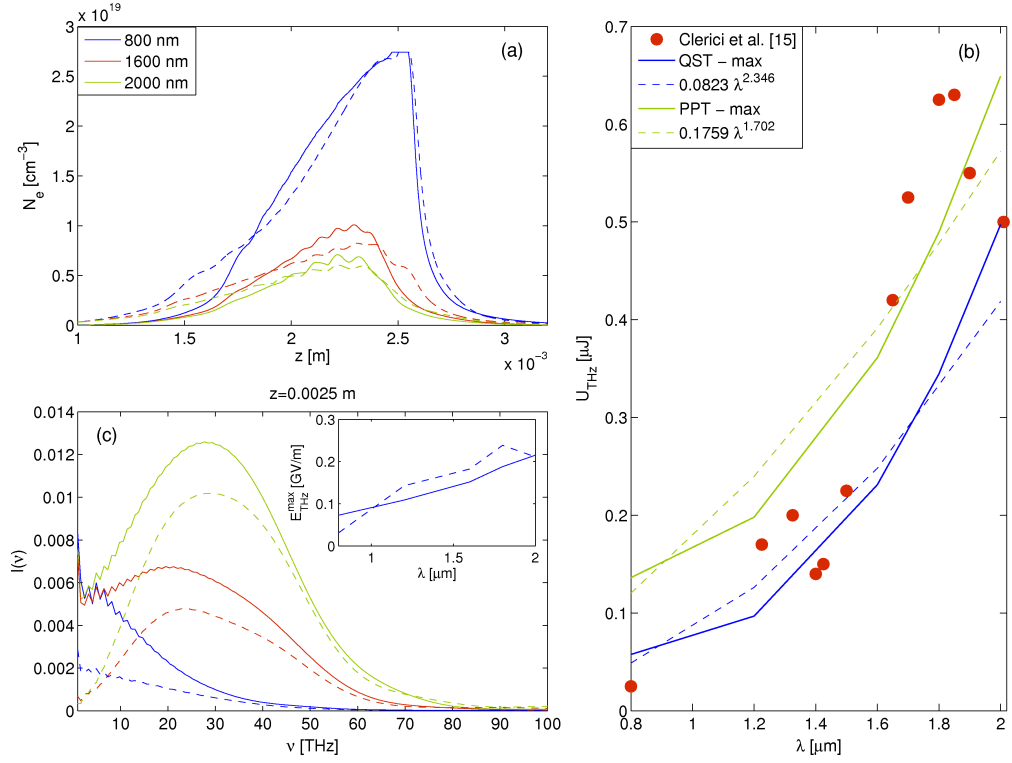}
\caption{Focused two-color Gaussian pulses in ratio $r = 5.2\%$ for QST ionization (dashed curves) and PPT ionization (solid curves) of O$_2$ and N$_2$ molecules. (a) Peak electron density. The selected pump wavelengths are 0.8, 1.6 and 2 $\mu$m (see legend). (b) Maximum THz energy yield for $\nu \leq 80$ THz as a function of the pump wavelength for a focused beam and two species ionized with the QST rate (blue curves) and with an instantaneous PPT rate (green curves). Scaling curves in $\lambda^{\alpha}$ shown as dashed curves are evaluated through least-square fitting (see legend). The solid red circles recall the experimental data points of Ref. \cite{Clerici:prl:110:253901}. (c) THz spectra computed at focus. Inset shows maximum THz fields.}
\label{Fig13}
\end{figure}

We first choose the same $f$-number $\sim 14$ as in Clerici {\it et al.}'s experiments \cite{Clerici:prl:110:253901}. For numerical reasons we limit the initial beam width to $w_0 = 150$ $\mu$m for a focal length $f=2.5$ mm. Our two-color pulses have 400 $\mu$J in energy with $5.2\,\%$ being injected into the second harmonic. These simulations include both Kerr and Raman nonlinearities and our selected THz window is again 80 THz. The simulations use a temporal step $\Delta t = 75$ attoseconds and a transverse resolution of $\Delta x = \Delta y = 0.88$ $\mu$m. For completeness we also included ionization of nitrogen molecules using either a QST rate ($U_i = 15.6$ eV, $Z^* = 1$) or a field-dependent PPT rate for two species, adopting Talebpour {\it  et al.}'s charge numbers $Z_{O_2}^* = 0.53,\, Z_{N_2}^* = 0.9$ \cite{Talebpour:oc:163:29}. When using a QST rate for both O$_2$ and N$_2$ molecules, the peak intensity and electron density reach 650 TW/cm$^2$ and $2.7\times10^{19}$ cm$^{-3}$ (complete single ionization) near focus with a 800-nm pump, respectively [see Fig. \ref{Fig13}(a)]. With the PPT rate, the effective charge numbers being less than unity promote weaker ionization rates \cite{Nuter:josab:23:874}, which increases the maximum pulse intensity at focus. Again a complete single ionization of both molecular species is achieved at 800 nm.\\

Figure \ref{Fig13}(b) compares the corresponding THz energy yields obtained from either a QST or an instantaneous PPT rate with both oxygen and nitrogen. Except at 800 nm with the PPT rate, the computed THz energy growth and values appear in good {\it quantitative} agreement with Clerici {\it et al.}'s experimental results [compare solid curves and red dots of Fig. \ref{Fig13}(b)]. Differences due to the choice of the ionization model are limited. A $\lambda$-dependent scaling of the THz yield appears closer to $\lambda^2$ than $\lambda^4$. Figure \ref{Fig13}(c) displays the spectra at focus. It is interesting to observe that the computed THz spectra now shifts their maximum to $\nu \simeq 30$ THz,  and not to $\nu = 5$ THz as reported in Ref. \cite{Clerici:prl:110:253901}. This discrepancy may be attributed to the fact that the ABCD technique used in \cite{Clerici:prl:110:253901} can barely measure frequencies above $\sim 20$ THz (see also \cite{Andreeva:prl:116:063902}). Inset details the growth of the maximum THz electric field, which increases quasi-linearly with the pump wavelength. Comparing Figs. \ref{Fig7a}(c) and \ref{Fig13}(b) demonstrates how sensitive the gain curves and THz spectra can be when one varies the laser parameters in focusing geometries.\\

\begin{figure}
\centering \includegraphics[width=\columnwidth]{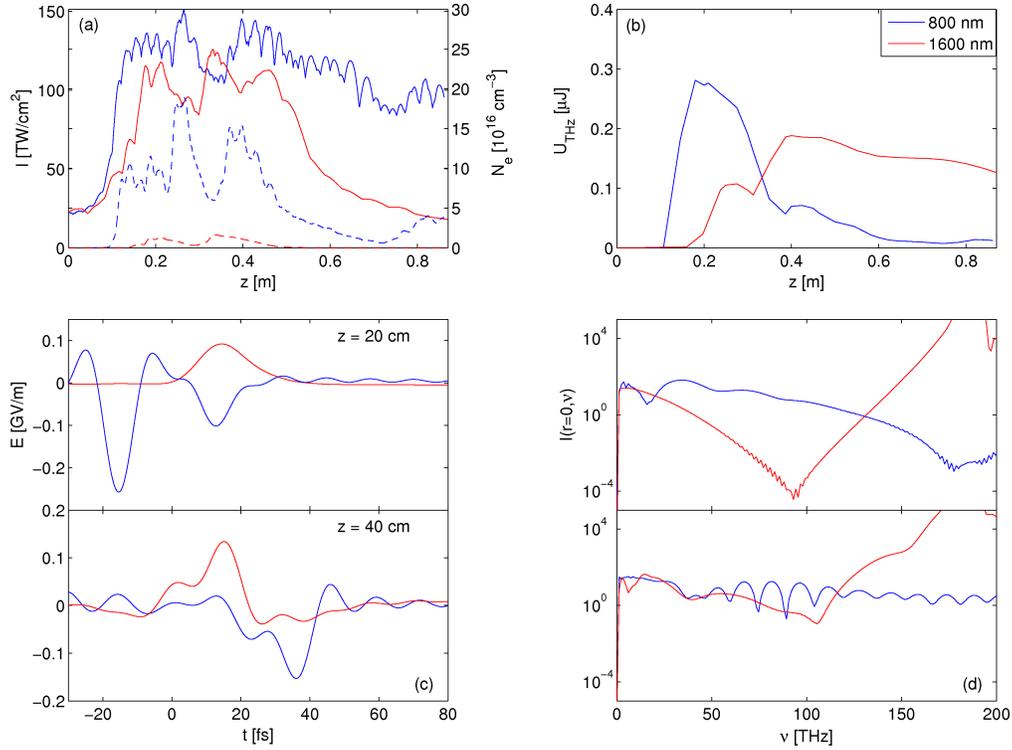}
\caption{Two-color filaments simulated with the Kerr indices and fractions of delayed nonlinearity reported in \cite{Wahlstrand:pra:85:043820,Zahedpour:ol:40:5794} for 800-nm (blue curves) and 1600-nm (red curves) pump pulses. (a) Peak intensities (LHS axis, solid curves) and maximum plasma densities (RHS axis, dashed curves). (b) THz energy in a 80-THz frequency window. (c) On-axis THz fields at $z=20$ cm and $z = 40$ cm, and (d) corresponding spectra.}
\label{Fig14}
\end{figure}

Finally, to evaluate the influence of the nonlinearity coefficients, we present in Fig. \ref{Fig14} the peak intensity and plasma density, THz energy, spectra and fields of the same femtosecond pulses as in Fig. \ref{Fig8} subject to stronger self-focusing with higher Raman-delayed responses, $n_2 = 3.79\times 10^{-19}$ cm$^2$/W, $x_k = 0.79$ as recently measured in \cite{Wahlstrand:pra:85:043820,Rosenthal:jpb:48:094011} for 800-nm pump pulses. With 1600-nm pump pulses, following Ref. \cite{Zahedpour:ol:40:5794}, we selected the Kerr index values $n_2 = 3.72\times 10^{-19}$ cm$^2$/W, $x_k = 0.78$. For completeness, we employed an instantaneous PPT rate with effective charge numbers $Z_{O_2} = 0.53$ and $Z_{N_2} = 0.9$ \cite{Talebpour:oc:163:29}. Unlike in Fig. \ref{Fig8}, the 1600-nm pump is here able to trigger a self-focusing sequence, starting with an input power ratio over critical equal to 3.1 that corresponds to an effective power ratio of about 2.5. Figure \ref{Fig14}(b) shows the THz energy evolving with the propagation distance. Over the plasma zone $(0.1 \leq z \leq 0.7$ m), the laser-to-THz conversion efficiency with a 1600-nm pump wave compared to a 800-nm pump is smaller than in the previous filamentary configuration [see green symbols $+$  in Fig. \ref{Fig7a}(c)], which can be attributed to the larger drop in the peak plasma densities reached in Fig. \ref{Fig14}(a). The maximum THz yield achieved with the 1600-nm pump is smaller than that reached in Fig. \ref{Fig8}. Figures \ref{Fig14}(c,d) detail the THz fields and spectra at the distances of maximum THz energy emission. Computed for low frequencies $\leq 80$ THz, on-axis THz fields created with the 800-nm pump prevail. On the whole, the main trend reported about Fig.~\ref{Fig8} is retrieved: Compared to the THz gain factors achieved in focused geometry, doubling the pump wavelength for two-color filaments propagating over meter-range distances in air may not significantly increase the THz conversion efficiency.\\

\section{Conclusion}

In summary, we have theoretically studied the influence of long pump wavelengths belonging to the range 0.8-2 $\mu$m in THz emissions caused by two-color laser pulses through air photoionization. We also cleared up the action of Raman-delayed and instantaneous Kerr nonlinearities of air molecules on the laser-to-THz conversion efficiency. Optical nonlinearities contribute to THz generation, in particular prior to ionization, but rotational Raman scattering leads to weaken the THz energy yield. At clamping intensity, direct laser-to-THz conversion via four-wave-mixing is weak compared to the photocurrent mechanism. However, Kerr-induced propagation effects such as cross-phase modulation and self-steepening have significant impact on THz generation. Furthermore, increasing the pump wavelength can dramatically enhance the THz energy yield. We showed that the THz energy gain factor cannot be quantitatively formulated with a simple power law in $\lambda^{\alpha}$ due to the influence of the relative phase between the two colors and their pulse envelopes. However, powers of growth rates between 2 and 5 can be justified from the local current model, mainly depending on the relative phase between the two colors. Scalings in $\sim \lambda^{2-3.5}$ have been extracted in focused propagation geometries through 3D comprehensive UPPE simulations, which faithfully reproduce experimental measurements of THz pulse energies. Similar gain factors can, however, barely be reached in filamentation geometry that clamps the intensity at smaller values and features much lower peak plasma densities at longer wavelengths. These results should help anticipate the THz gain factors achieved with mid-infrared laser systems used in future experiments.

\section{Acknowledgements}

This work was performed using HPC resources from PRACE (Grant \# 2014112576) and GENCI (Grant \# 2016-057594). IB acknowledges support by the joint grant DFG-MO $\#$ 850/20-1 - RSF $\#$ 16-42-01060.

\end{document}